\input{aipcheck}

\documentclass[
    ,final            
  ]
  {aipproc}

\layoutstyle{6x9}

\def\Im{\mathrm {Im}}
\def\Re{\mathrm {Re}}

\def\degree{\kern-.2em\r{}\kern-.3em}

\begin{document}

\title{Measurements of Single and Double Spin Asymmetry in \textit{pp} Elastic Scattering 
              in the CNI Region with Polarized Hydrogen Gas Jet Target}

\classification{13.88.+e, 13.85.Dz, 29.27.Pj, 29.27.Hj}

\keywords {Elastic scattering, spin, coulomb nuclear interference}

\author{H.~Okada}{
  address={Kyoto University, Kyoto Japan} 
  ,altaddress={RIKEN, Wako, JAPAN}
}

\author{I.~Alekseev}{
 address={Institute for Theoretical and Experimental Physics (ITEP), 117259 Moscow, Russia}
}
\author{A.~Bravar}{
 address={University of Geneva , 1205 Geneva, Switzerland}
}
\author{G.~Bunce}{
 address={Brookhaven National Laboratory, Upton, NY 11973, USA}
,altaddress={RIKEN BNL Research Center, Upton, NY 11973, USA} 
}
\author{S.~Dhawan}{
 address={Yale University, New Haven, CT 06520, USA}
}
\author{K.O.~Eyser}{
 address={University of California, Riverside, CA 92521, USA}
}
\author{R.~Gill}{
 address={Brookhaven National Laboratory, Upton, NY 11973, USA}
}
\author{W.~Haeberli}{
 address={University of Wisconsin, Madison, WI 53706, USA}
}
\author{H.~Huang}{
 address={Brookhaven National Laboratory, Upton, NY 11973, USA}
}
\author{O.~Jinnouchi}{
 address={KEK, Tukuba, Japan}
}
\author{Y.~Makdishi}{
 address={Brookhaven National Laboratory, Upton, NY 11973, USA}
}
\author{I.~Nakagawa}{
 address={RIKEN, Wako, JAPAN }
}
\author{A.~Nass}{
 address={University of Erlangen, 91058 Erlangen, Germany}
}
\author{N.~Saito}{
 address={Kyoto University, Kyoto Japan} 
 ,altaddress={KEK, Tukuba, Japan}
}
\author{E.~Stephenson}{
 address={Indiana University Cyclotron Facility, Bloomington, IN 47408, USA}
}
\author{D.~Sviridia}{
 address={Institute for Theoretical and Experimental Physics (ITEP), 117259 Moscow, Russia}
}
\author{T.~Wise}{
 address={University of Wisconsin, Madison, WI 53706, USA}
}
\author{J.~Wood}{
 address={Brookhaven National Laboratory, Upton, NY 11973, USA}
}
\author{A.~Zelenski}{
 address={Brookhaven National Laboratory, Upton, NY 11973, USA}
}

\begin{abstract}
 Precise measurements of the single spin asymmetry, $A_N$ and the double spin asymmetry, $A_{NN}$, in proton-proton (\textit{pp}) elastic~scattering~in~the~region
of four-momentum transfer squared $0.001 < -t < 0.032~({\rm GeV}/c)^2$ have been performed
using a polarized atomic hydrogen gas jet target and the RHIC polarized proton beam at $24$~GeV/$c$ and $100$~GeV/$c$.
The polarized gaseous proton target allowed us to achieve the measurement of $A_{NN}$ in the CNI region for the first time.  
Our results of $A_N$ and $A_{NN}$ provide significant constraints to determine the magnitude
of poorly known hadronic single and double spin-flip amplitudes at this energy.
\end{abstract}

\maketitle


\paragraph{Introduction}
	\textit{pp} elastic scattering is one of the most fundamental reactions in particle-nuclear physics 
	and is described in transition amplitudes by use of helicity of initial and final states.
	Requiring that the interaction is invariant under space inversion, time reversal and 
        rotation in spin space, \textit{pp} scattering in a given
	spin state is described in five independent transition amplitudes ($\phi_i,~i=1-5$) as functions of 
        the center-of-mass energy squared,~$s$, and $t$~\cite{WickYacob}.
	The understanding of  these amplitudes would provide crucial guidelines to investigate the reaction mechanism.

\begin{figure}
\begin{tabular}{cc}
\begin{minipage}{0.5\hsize}
\includegraphics[height=0.25\textheight,angle=-90]{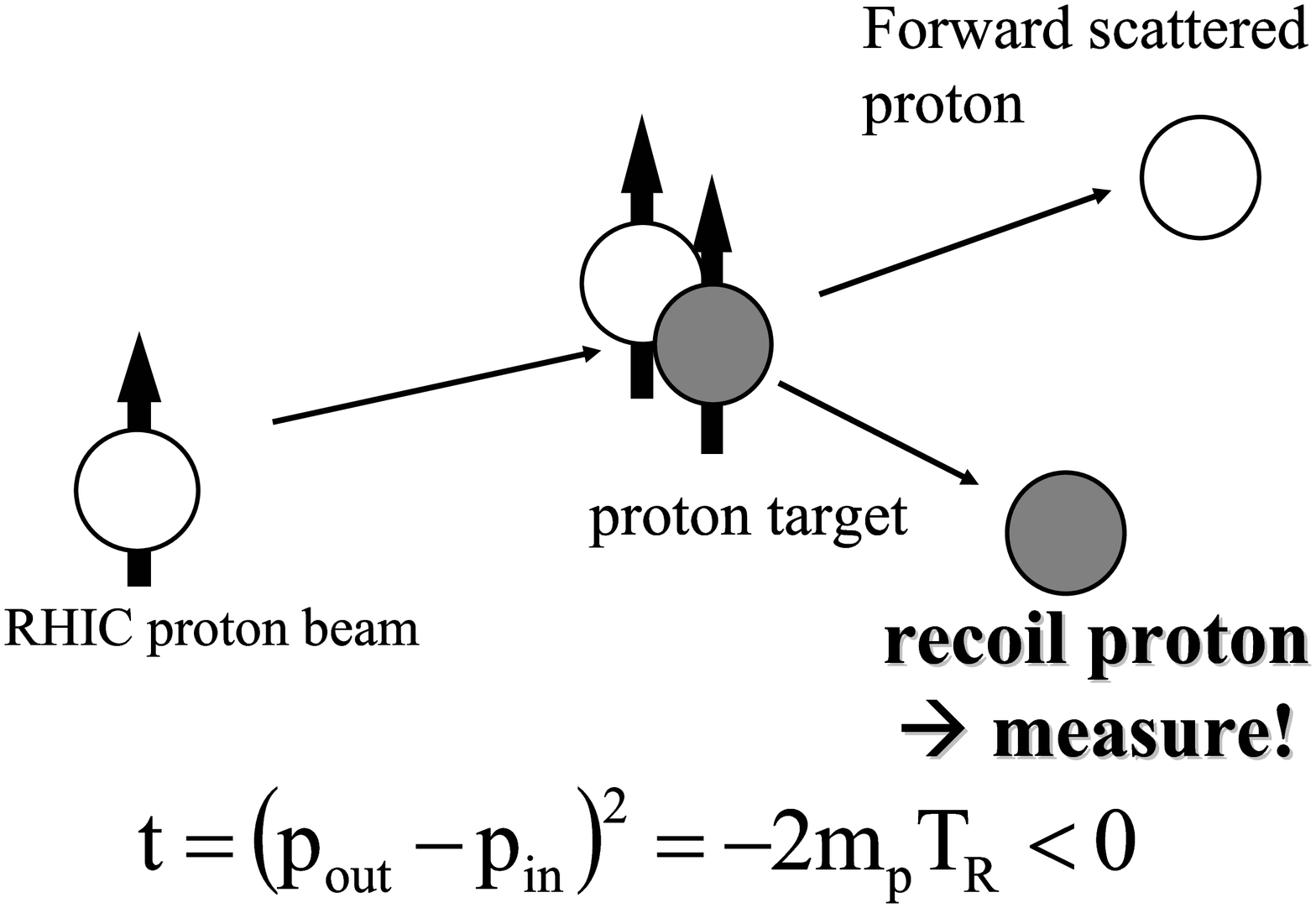}
\caption{aaa
}
\label{fig1-2}
\end{minipage}
\begin{minipage}{0.5\hsize}
\includegraphics[height=0.25\textheight]{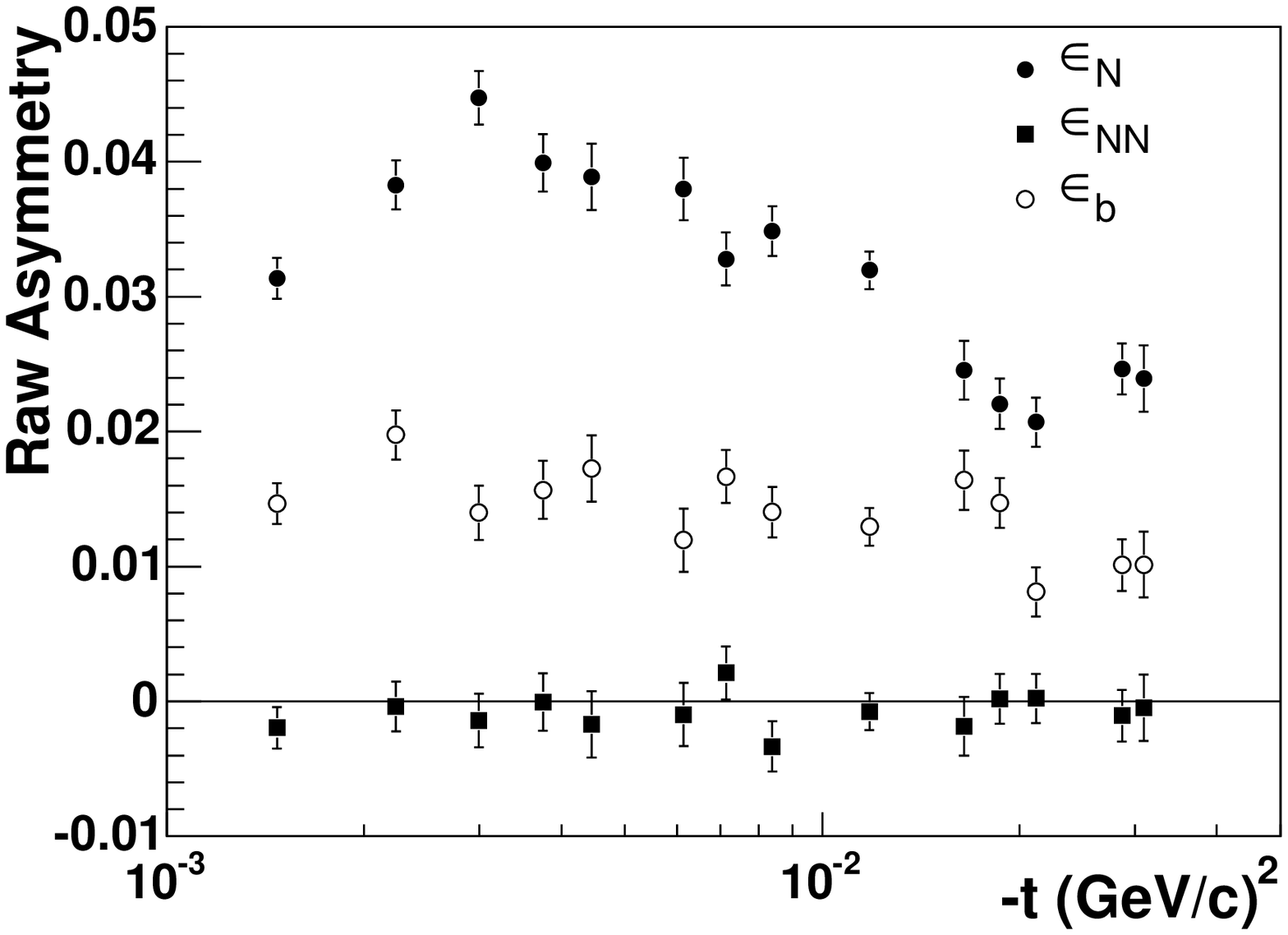}
\caption{Left: Example of "parallel" case, $p^{\uparrow}p^{\uparrow} \rightarrow pp$. 
         Right: $\epsilon_{NN}$, $\epsilon_N$ and $\epsilon_{b}$ at $\sqrt{s}=13.7$~GeV as a function of $-t$.
}
\label{rawAsymmetries}
\end{minipage}
\end{tabular}
\end{figure}

	Each transition amplitude is described as a sum of the hadronic amplitude ($\phi_i^{had}$) and 
        the electro-magnetic amplitude ($\phi_i^{em}$).
	In the small $-t$ region, $\phi_i^{em}$ and $\phi_i^{had}$ become similar in strength 
        and interfere with each other. 
        We call this interference the Coulomb Nuclear Interference (CNI).
	Thanks to the great successes of QED and the past precisely measured quantities (ex. magnetic moment), 
        $\phi_i^{em}$ is precisely described.
	On the other hand, $\phi_i^{had}$ is not fully described by theory, because the perturbative QCD 
	is not applicable in the CNI region.
	By use of the experimental data of total and differential cross-section of unpolarized \textit{pp} elastic 
        scattering, we can determine the sum of two non-spin-flip hadronic amplitudes 
        ($\phi_+^{had}=\phi_1^{had}+\phi_3^{had}$)~\cite{UnpolExp}.
	In order to approach to the hadronic single and double spin-flip amplitudes ($\phi_5^{had}$ and $\phi_2^{had}$), 
	we measure $A_N$ and $A_{NN}$ in the CNI region.
        $A_N$ is defined by the asymmetry of cross-section with up-down transverse polarization for one of the protons.
        Similarly $A_{NN}$ is defined by the asymmetry of cross-section for parallel and anti-parallel 
        transverse polarization for both of the protons. 
	The left side of Fig.~\ref{rawAsymmetries} depicts "parallel" case. We define the scattering plane 
        from $3$-momenta of incident and recoil particles, which is normal to the spin directions.
        By use of transition amplitudes, $A_N$ is expressed as,
	\begin{equation}
	  A_N \approx  \frac{-\Im[ \phi_5^{em}(s,t)\phi_+^{had*}(s,t)+\phi_5^{had}(s,t)\phi_+^{em}(s,t)]}
                            {|\phi_+(s,t)|^2}.
		\label{eq:AN_helicity} 
	\end{equation}
	The first term of Eq.~\ref{eq:AN_helicity} is calculable and has a peak around 
        $-t \simeq 0.003 ~({\rm GeV}/c)^2$~\cite{But99} which is generated by proton's anomalous magnetic moment. 
	Because the presence of $\phi_5^{had}$ introduces a deviation in shape and magnitude from the first term, 
	a measurement of $A_N$ in the CNI region, therefore, can be a sensitive probe for $\phi_5^{had}$.

	$A_{NN}$ is expressed as,
	 \begin{equation}
		A_{NN}\approx \frac{2|\phi_5^{had}(s,t)|^2+\Re[(\phi_+(s,t))^{*}\phi_2^{had}(s,t)]}{|\phi_+(s,t)|^2}.
          \label{eq:ANN_helicity} 
        \end{equation}
	Because the first term is $2$nd order of $\phi_5^{had}$ and the second term is $1$st order of $\phi_2^{had}$, 
        $A_{NN}$ is sensitive to $\phi_2^{had}$~\cite{ANN_New}.
        From a consequence of angular momentum conservation at small $-t$ and large $\sqrt{s}$,
	we use $\phi_4^{had} \propto~t~\rightarrow 0$ for these expressions.

\paragraph{Experiment}
	Experiment has been performed using a polarized hydrogen gas jet target and polarized RHIC proton beam at 
        $24$~GeV/$c$ ($\sqrt{s}=6.7$ GeV) and $100$~GeV/$c$ ($\sqrt{s}=13.7$ GeV).
        We detect the recoil protons by silicon detectors which are located on both sides of the target.
	The details of experimental setup are described in~\cite{hiromi06,Zel05}.
	
        In the \textit{pp} elastic scattering process, both forward-scattered particle and 
        recoil particle are protons and there are no other particles 
	involved nor new particles produced in the process. 
	Since initial states are well defined, the elastic process can be, in principle, 
        identified by detecting the recoil particle only.
	By measuring kinetic energy $T_R$, time of flight, and recoil angle of recoil particle, 
        we measure the mass of recoil particle and all the forward scattered rest particles.
	We collected $4.3$~M events at $\sqrt{s}=13.7$~GeV and $0.8$~M events at $\sqrt{s}=6.7$~GeV, respectively. 
	The details of event selection is described in ~\cite{hiromi06}.

	The selected event yield is sorted by $-t$ bins, which is obtained measuring the kinetic energy of 
        the recoil particle:~$-t= 2m_pT_R$, and spin states. $m_p$ is the proton mass.
	Then we calculate two types of single spin \textit{raw} asymmetries, $\epsilon_N$ for the target spin state and 
        $\epsilon_{b}$ for the beam spin state. 
	We also calculate double spin \textit{raw} asymmetry, $\epsilon_{NN}$ for the target and beam spin states. 

	The right side of Fig.~\ref{rawAsymmetries} displays these \textit{raw} asymmetries 
        of $\sqrt{s}=13.7$~GeV data as a function of $-t$ in the region $0.001~\leq-t~\leq~0.035$~(GeV/$c$)$^2$  
	($0.5~\leq~T_R~\leq~17$~MeV).
        The polarized gaseous proton target allowed us to achieve the measurement in the CNI region 
        for the first time.  
	In order to cancel out the asymmetries of up-down luminosity, and detector acceptance, 
        we employed so-called "square root formula" for $\epsilon_N$ and $\epsilon_b$ calculation. 
	On the other hand, $\epsilon_{NN}$ needs to be corrected by the luminosity asymmetry. 
        The target spin flips every $5$ minutes and the density of both spin states are 
        the same and stable during the experimental period ($\sim90$ hours). 
        The beam intensity , which is measured by the wall current monitor~\cite{WCM}, varies by bunch (every $106$~nsec in $2004$) 
        and fill (every several hours). 
        By accumulating intensity for the experimental period, the variation is compensated. 
	Therefore the luminosity asymmetry is quite small compared to statistical error of $\epsilon_{NN}$.

\paragraph{Results and discussions}
        $A_N$ is measured normalizing $\epsilon_N$ by well measured the target polarization $P_t$~\cite{Zel05},
	\begin{equation}
		A_N = \frac{\epsilon_N}{P_t}.
	\end{equation}
	
	Utilizing the measured $A_N$, we also measure the beam polarization,~$P_b=\epsilon_b/A_N$ 
        \footnote{This experimental setup also plays an important role in the RHIC spin 
                  program to measure the absolute beam polarization~\cite{Oleg}.}.
        Normalizing $\epsilon_{NN}$ by $P_t$ and $P_b$, $A_{NN}$ is obtained via
	\begin{equation}
		A_{NN}=\frac{\epsilon_{NN}}{P_bP_t}. 
	\end{equation}

\begin{figure}
\begin{tabular}{cc}
\begin{minipage}{0.5\hsize}
\includegraphics[height=0.25\textheight]{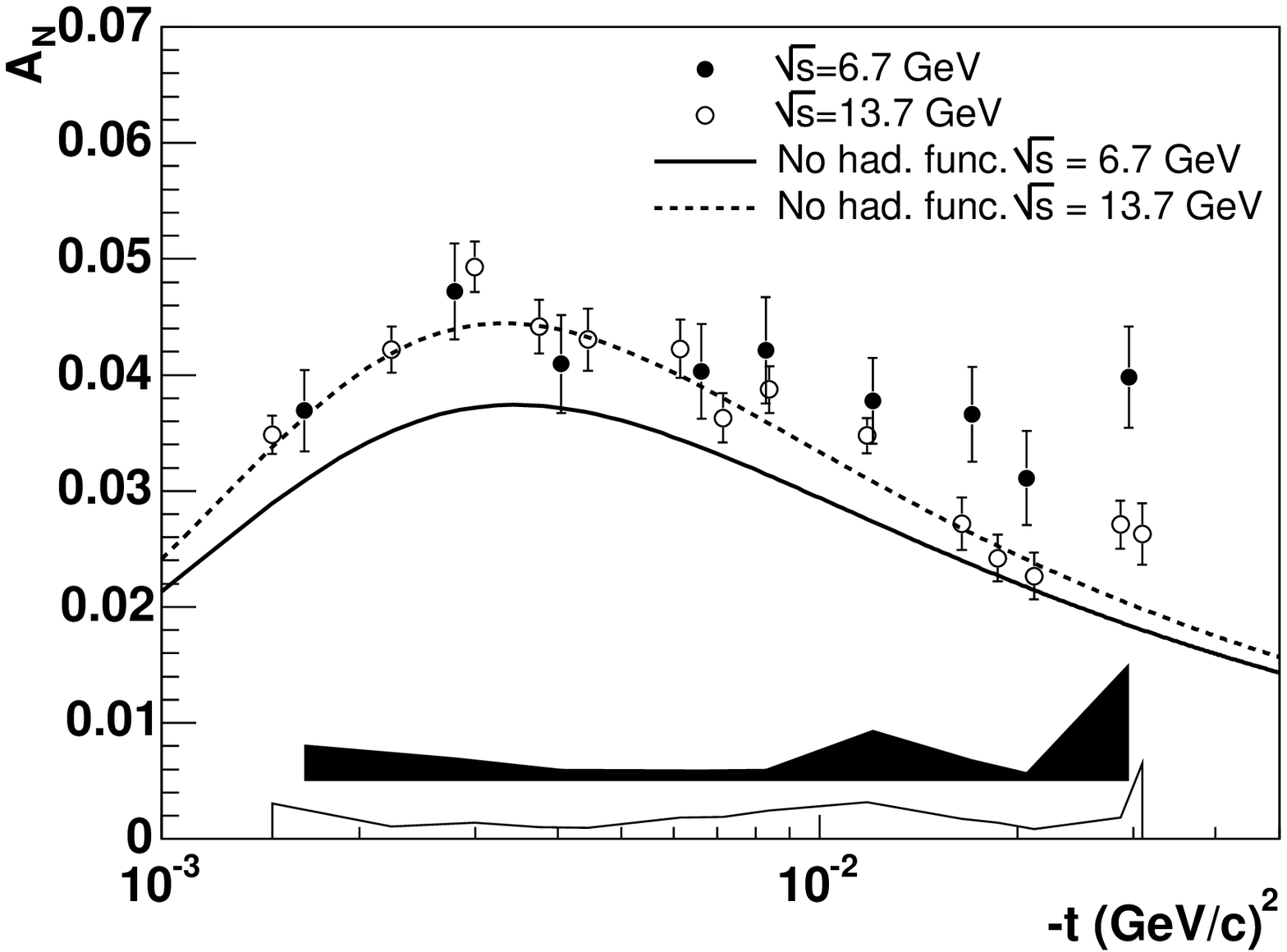}
\label{AN}
\end{minipage}
\begin{minipage}{0.5\hsize}
\includegraphics[height=0.25\textheight]{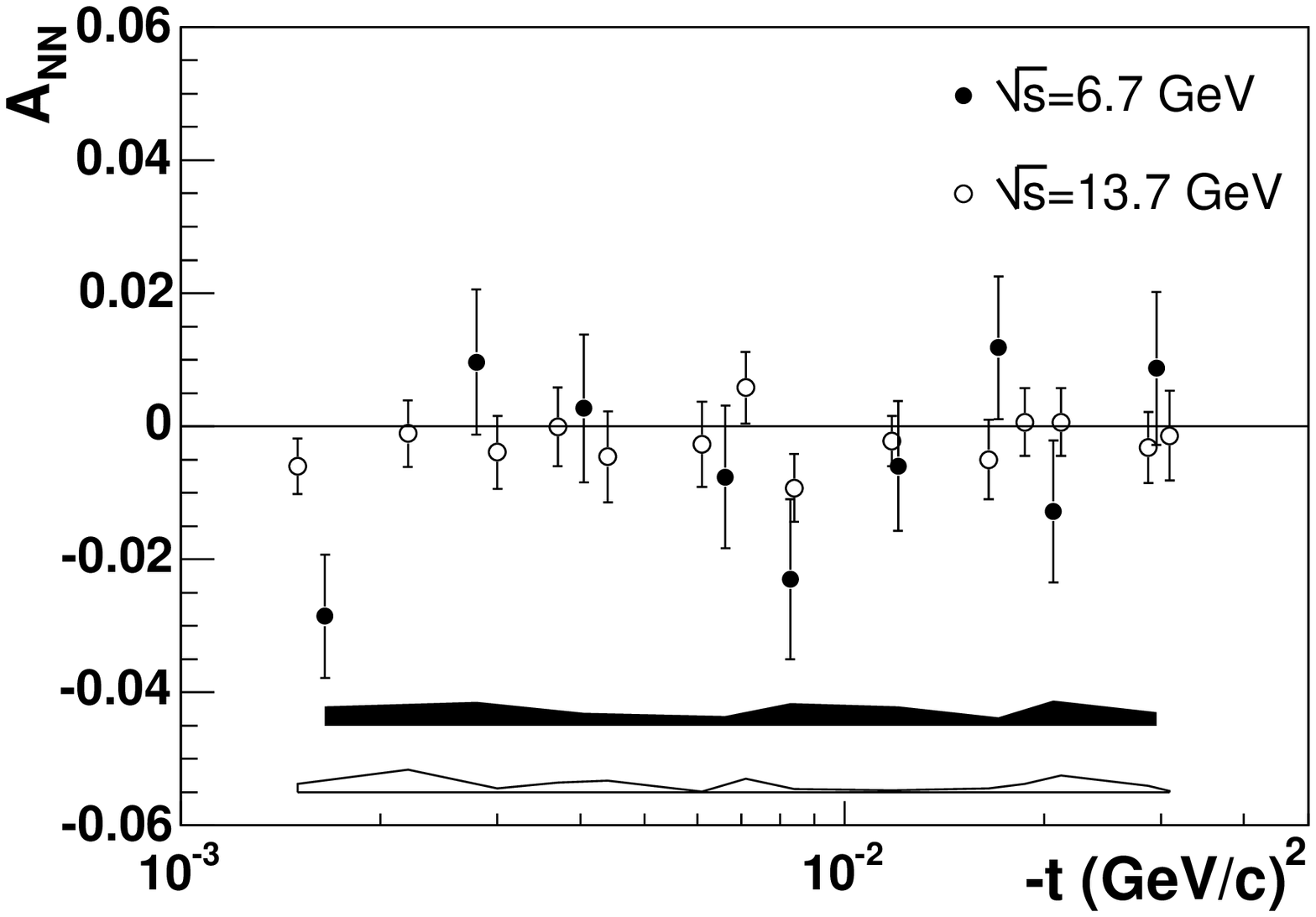}
\caption{The left and right plots display the results of $A_N$ and $A_{NN}$ at $\sqrt{s}=6.7$~GeV 
        with filled circles and $13.7$~GeV with open circles, respectively.
	The errors on the data points are statistical. The lower bands represents the total systematic errors.
	The solid and dashed lines in the left plots correspond to the first term in Eq.~\ref{eq:AN_helicity} 
        for these $\sqrt{s}$, respectively. 
}
\label{ANN}
\end{minipage}
\end{tabular}
\end{figure}

	The left and right plots of Fig.~\ref{ANN} display the results of $A_N$ and $A_{NN}$ at $\sqrt{s}=6.7$~GeV 
        with filled circles and $13.7$~GeV with open circles, respectively.
	The errors on the data points are statistical. The lower bands represent the total systematic errors.
	The solid and dashed lines correspond to the first term in Eq.~\ref{eq:AN_helicity} 
        for these $\sqrt{s}$, respectively. 

	$A_N$ at $\sqrt{s}=13.7$~GeV are consistent with the dashed line ($\chi^2$/ndf=$13.4/14$). 
        On the other hand, although the accuracy is statistically limited, 
        $A_N$ at $\sqrt{s}=6.7$~GeV are \textit{not} consistent with the solid line 
        ($\chi^2$/ndf=$35.5/9$) and this discrepancy implies the presence of $\phi_5^{had}$.
	$A_{NN}$ for these $\sqrt{s}$ have no clear $-t$ dependence and the average values are consistent 
        with zero within $1.5~\sigma$.

	In summary, measurements of $A_N$ and $A_{NN}$ provide experimental knowledge to poorly known 
        $\phi_5^{had}$ and $\phi_2^{had}$. 
	The $\sqrt{s}$ dependence of $\phi_5^{had}$ is provided by $A_N$ results and the theoretical interpretation 
        is under way~\cite{Larry}.  
	However, there is no comprehensive understanding of $\phi_2^{had}$ and $\phi_5^{had}$ yet.
        Further measurements at different $\sqrt{s}$ are required to fully describe the behavior 
        of $\phi_2^{had}$ and $\phi_5^{had}$. 




\bibliographystyle{aipprocl} 






\end{document}